\documentclass[conference,10pt]{IEEEtran}
\IEEEoverridecommandlockouts

\usepackage{amsthm,amssymb}
\usepackage{graphicx}
\usepackage{braket}
\usepackage[dvipsnames]{xcolor}
\usepackage{arydshln}
\usepackage[utf8]{inputenc} 
\usepackage[T1]{fontenc}
\usepackage{url}
\usepackage{ifthen}
\usepackage{cite}
\usepackage[cmex10]{amsmath} 
\usepackage{bm}
\usepackage{amsfonts}
\usepackage{xcolor}
\usepackage{bbm}
\usepackage{colortbl}
\usepackage{enumitem}

\newtheorem{theorem}{Theorem}

\newtheorem{corollary}{Corollary}

\newtheorem{remark}{Remark}

\newtheorem{example}{Example}

\newcommand{\defeq}{\stackrel{\Delta}{=}}

\begin{document}



\title{Breaking the Storage-Bandwidth Tradeoff in Distributed Storage with Quantum Entanglement}

\author{Lei Hu \quad Mohamed Nomeir \quad Alptug Aytekin \quad Sennur Ulukus\\
\normalsize Department of Electrical and Computer Engineering\\
\normalsize University of Maryland, College Park, MD 20742 \\
\normalsize \emph{leihu@umd.edu} \quad \emph{mnomeir@umd.edu} \quad \emph{aaytekin@umd.edu}  \quad \emph{ulukus@umd.edu}}

\maketitle

\begin{abstract}
    This work investigates the use of quantum resources in distributed storage systems. Consider an $(n,k,d)$ distributed storage system in which a file is stored across $n$ nodes such that any $k$ nodes suffice to reconstruct the file. When a node fails, any $d$ helper nodes transmit information to a newcomer to rebuild the system. In contrast to the classical repair, where helper nodes transmit classical bits, we allow them to send classical information over quantum channels to the newcomer. The newcomer then generates its storage by performing appropriate measurements on the received quantum states. In this setting, we fully characterize the fundamental tradeoff between storage and repair bandwidth (total communication cost). Compared to classical systems, the optimal storage--bandwidth tradeoff can be significantly improved with the enhancement of quantum entanglement shared only among the surviving nodes, particularly at the minimum-storage regenerating point. Remarkably, we show that when $d \geq 2k-2$, there exists an operating point at which \textit{both storage and repair bandwidth are simultaneously minimized}. This phenomenon breaks the tradeoff in the classical setting and reveals a fundamentally new regime enabled by quantum communication.
\end{abstract}

\section{Introduction}
Distributed storage systems are a fundamental infrastructure for modern data centers and cloud storage platforms \cite{Rhea_storage, Bhagwan_storage, Kamra_storage}. Since individual storage nodes may be unreliable, a file is encoded and stored across $n$ nodes such that the original file can be reconstructed by accessing the contents of any $k$ nodes (referred to as the \textit{data retrieval} property) \cite{Dimakis_2010, Patra_2025}. To maintain this property in the presence of node failures, the system initiates a repair process whenever an individual storage node fails. Specifically, a newcomer node connects to any $d$ surviving helper nodes, each of which transmits $\beta$ dits of information to the newcomer. Based on the received information, the newcomer generates its stored content so that the data retrieval property of the system is preserved. To describe the performance limit of the system, Dimakis \emph{et al.}~\cite{Dimakis_2010} characterized the fundamental tradeoff between per-node storage and repair bandwidth, showing that these two resources cannot be simultaneously minimized in general and thus there is a tradeoff, i.e., reducing storage necessarily increases repair bandwidth, and vice versa.

Over the past decades, quantum technology has been shown to enhance the efficiency of classical communication significantly \cite{shi2021entanglement, nielsen2010quantum, hsieh2008entanglement}. A typical example is superdense coding, in which a single qubit, when assisted by pre-shared entanglement between the transmitter and the receiver, can convey up to two classical bits of information. More recently, quantum resources have also demonstrated significant advantages in distributed computing scenarios \cite{yao_capacity_MAC, aytekin2023quantum, entanglement_assisted, yao2024inverted}, where a user seeks to compute a function of classical data stored across multiple servers. In such settings, shared entanglement only among the servers (i.e., transmitters) can substantially reduce the required communication cost compared to its classical counterpart. These observations naturally raise the question of whether quantum communication can be exploited to improve the fundamental performance limits of distributed storage systems. In particular, we seek to characterize the new storage-bandwidth tradeoff when helper nodes are allowed to transmit quantum states (qudits), and to determine the maximum performance gains achievable via entanglement assistance among the surviving nodes (transmitters).

Motivated by this question, we investigate the performance gains enabled by quantum communication in distributed storage systems. We consider an entanglement-assisted repair model in which the helper nodes share a pre-established entangled quantum state and locally encode their stored classical information into their states, which are then transmitted to the newcomer. Upon receiving these quantum states, the newcomer performs appropriate measurements to generate its stored content, while ensuring that the resulting storage system continues to satisfy the data retrieval property.

Within this framework, we fully characterize the fundamental tradeoff between storage and repair bandwidth. The detailed converse and achievability proofs will be presented in \cite{QRC_arxiv}, due to space limitations here. In this paper, we focus on analyzing the performance gains enabled by quantum resources at both the minimum storage regeneration (MSR) point and the minimum bandwidth regeneration (MBR) point, and provide a pictorial explanation to explicitly show the quantum advantage regions. Our results show that when $d \leq 2k-2$, entanglement-assisted repair can reduce the required repair bandwidth by a factor of two at the MSR point. Further, when $d \geq 2k-2$, we identify a unique operating point at which both the repair bandwidth and per-node storage are simultaneously minimized, which does not occur in classical systems.

\textit{Notation:}
For an integer $M$, $[M]$ denotes the set $\{1, \ldots, M\}$. For integers $a$ and $b$, $[a:b]$ denotes  $\{a,\ldots,b \}$. $A_{[M]}$ is the compact notation of $\{A_{1}, A_2, \ldots, A_{M}\}$. $\otimes$ denotes the tensor product.

\section{Problem Statement \& Classical Results}
\subsection{System Model}
We consider an $(n,k,d)$ distributed storage system storing a (classical) file consisting of $B$ dits across $n$ storage nodes, where each node stores $\alpha$ dits. The system is required to satisfy the following properties:
\begin{itemize}
    \item \textbf{(Data retrieval)} The entire file can be reconstructed from the contents of any $k$ nodes;
    \item \textbf{(Single-node repair)} Upon the failure of a storage node, a newcomer node connects to any $d$ surviving nodes among the remaining $n-1$ nodes, where $k \leq d \leq n-1$. After the repair process, the resulting system must continue to satisfy the data retrieval property.
\end{itemize}
We consider the repair process under classical and quantum communication models specified as follows.

\subsubsection{Classical Repair}
In the classical setting, each of the $d$ surviving nodes transmits $\beta_{\mathsf{c}}$ dits to the newcomer, resulting in a total repair communication cost (repair bandwidth) of $d\beta_{\mathsf{c}}$. A fundamental result for this $(n,k,d,\alpha,\beta_{\mathsf{c}},B)$ system is that there exists a tradeoff between the per-node storage $\alpha$ and the repair bandwidth $d\beta_{\mathsf{c}}$: increasing storage can reduce the required repair bandwidth, and vice versa. This storage–bandwidth tradeoff was fully characterized in \cite{Dimakis_2010}, and is reviewed in Section~\ref{sec:classical_res}.

\subsubsection{Quantum Repair}
In this paper, we explore performance improvement by allowing \textit{quantum communication} during the repair process. Specifically, the $d$ helper nodes are assumed to share a prior entangled state. Each helper node transmits $\beta_{\mathsf{q}}$ qudits to the newcomer. By performing appropriate quantum measurements on the received quantum systems, the newcomer reconstructs $\alpha$ classical dits to be stored locally, such that the repaired system continues to satisfy the \textit{data-retrieval} property.

The quantum communication-assisted repair steps are specified as follows.
\begin{enumerate}[label=\alph*)]
    \item \textbf{(Entanglement distribution)}
    Suppose the newcomer node communicates with $d$ nodes among the $n-1$ surviving nodes, indexed by $s_1,\ldots,s_d$. These $d$ nodes share a globally entangled quantum state represented by the density matrix $\rho^{\rm ini}$ in the quantum system $\mathcal{Q} = \mathcal{Q}_1 \cdots \mathcal{Q}_d$. This state is pre-established before the communication begins. The dimension of each quantum subsystem $\mathcal{Q}_s$ is given by $\delta_s \defeq|\mathcal{Q}_s|$, and each subsystem consists of $\beta_{\mathsf{q}}$ qudits, i.e., $\log_q \delta_s = \beta_{\mathsf{q}} $ where $q$ is a prime power.
    \item \textbf{(Information encoding)}
    Each node $s \in \{s_1,\ldots,s_d\}$ applies a local encoding operation to its quantum state on its subsystem $\mathcal{Q}_s$ in order to embed classical information determined by its stored data. Specifically, node $s$ applies a completely positive trace-preserving (CPTP) map $\Lambda_s$ to its quantum state based on its local storage content. After all encoding operations are applied, the joint quantum state is given by
    \begin{align}
        \rho^{\rm enc} =   \bigotimes_{s \in \{s_1,\ldots,s_d\} } \Lambda_s  (\rho^{\rm ini}).
    \end{align}
    The $d$ encoded quantum subsystems are then transmitted to the newcomer through separate \textit{noiseless} quantum channels.
    \item \textbf{(Storage generation)}
    Upon receiving the encoded quantum state, $\rho^{\rm enc} $, the newcomer performs a suitable positive-operator-valued measure (POVM). The resulting measurement outcomes allow the newcomer to generate $\alpha$ dits as its storage, such that the repaired system continues to satisfy the \textit{data retrieval} property.
\end{enumerate}

Our objective is to characterize the new tradeoff between storage and repair bandwidth for this $(n,k,d,\alpha,\beta_{\mathsf{q}},B)$ system. We compare the tradeoffs in the classical and quantum settings in order to show the potential quantum advantage in distributed storage systems. To this end, we next review the result in classical distributed storage systems.

\subsection{Classical Results}\label{sec:classical_res}
The fundamental storage-repair bandwidth (in dits) tradeoff was established in \cite{Dimakis_2010}. Specifically, by applying the cut-set bound from network coding, any classical distributed storage system with parameters $(n,k,d,\alpha,\beta_{\mathsf{c}},B)$ must satisfy
\begin{align}
    \mathrm{[Classical]}: \quad \sum_{i=0}^{k-1} \min\{(d-i) \beta_{\mathsf{c}}, \alpha \} \geq B.\label{eq:classical}
\end{align}
This inequality characterizes all achievable tradeoff between per-node storage $\alpha$ and repair bandwidth $d\beta_{\mathsf{c}}$.

Two extremal points on the tradeoff curve are of particular interest: the \textit{minimum storage regeneration} (MSR) point and the \textit{minimum bandwidth regeneration} (MBR) point, which are specified as follows.
\begin{itemize}
    \item The MSR point is obtained by first minimizing the per-node storage $\alpha$ and then minimizing the per-node bandwidth $\beta_{\mathsf{c}}$. The corresponding operating point is
    \begin{align}
        \left(\alpha^{\mathrm{MSR}},d \beta_{\mathsf{c}}^{\mathrm{MSR}} \right) = \left(\frac{B}{k},\frac{B}{k} \cdot \frac{d}{d-k+1} \right).
    \end{align}
    \item The MBR point is obtained by first minimizing the $\beta_{\mathsf{c}}$ and then minimizing $\alpha$. The resulting operating point is
    \begin{align}
        \left(\alpha^{\mathrm{MBR}},d \beta_{\mathsf{c}}^{\mathrm{MBR}}\right) = \left(\frac{2Bd}{2kd - k^2 + k},\frac{2Bd}{2kd - k^2 + k} \right).
    \end{align}
\end{itemize}

\begin{remark}
    Note that $\alpha^{\rm MBR} \geq \alpha^{\rm MSR}$ and the equality holds if and only if the file is replicated among all nodes $(k = 1)$. Hence, for any non-replicated storage system, i.e., $k \geq 2$, the MSR point and MBR point do not coincide in the classical setting. In the following, we show that with the assistance of quantum communication and entanglement, it is possible to simultaneously minimize both the storage and repair bandwidth (in qudits).
\end{remark}

\section{Main Results}

\begin{theorem}\label{thm_cutset}
    Consider a distributed storage system with parameters $(n,k,d,\alpha,\beta_{\mathsf{q}},B)$. The optimal tradeoff between storage and repair bandwidth (measured in qudits) is characterized by
    \begin{align}
        \mathrm{[Quantum]}: \quad \sum_{i=0}^{k-1} \min\{2(d-i) \beta_{\mathsf{q}}, d \beta_{\mathsf{q}}, \alpha \} \geq B. \label{eq:main}
    \end{align}
\end{theorem}

The converse and achievability proofs of Theorem~\ref{thm_cutset} will be presented in \cite{QRC_arxiv}. To provide intuition for Theorem~\ref{thm_cutset}, we have the following remark.

\begin{figure}[t]
    \centering
    \includegraphics[width=0.48\textwidth]{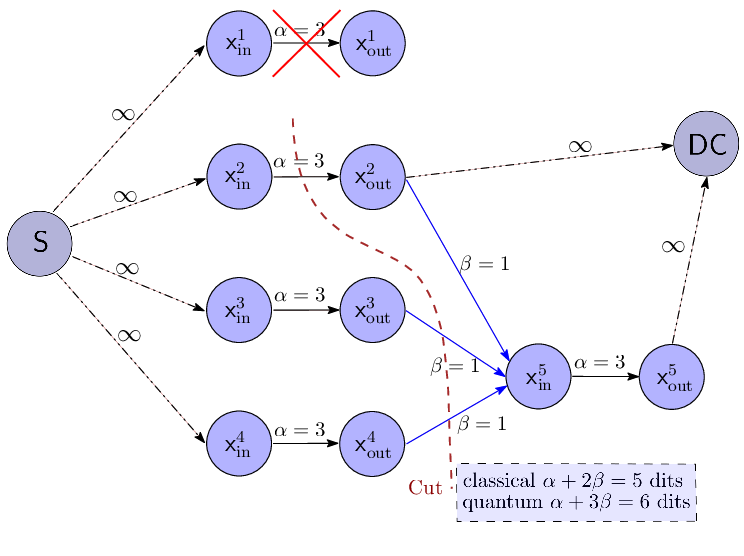}
    \caption{The classical and quantum cut for distributed storage system with $(n,k,d) = (4,2,3)$, $\alpha=3$, and $ \beta_{\mathsf{c}} = \beta_{\mathsf{q}} = \beta = 1$.}
    \label{fig:thm1}
\end{figure}

\begin{remark}
    The bound in Theorem~\ref{thm_cutset} can be interpreted via a cut-set argument on a quantum-communication-assisted information flow graph. To be specific, the information flow graph consists of a single data source $\mathsf{S}$, storage nodes $\mathsf{x}_{\rm in}^{i}, \mathsf{x}_{\rm out}^{i}$ for each $i \in [n]$, and a data collector $\mathsf{DC}$. The source node $\mathsf{S}$ represents the original file. Each storage node $i$ is modeled by a pair of vertices $\mathsf{x}_{\rm in}^{i}$ and $\mathsf{x}_{\rm out}^{i}$ connected by a directed edge with capacity equal to the storage size at node $i$. A data collector $\mathsf{DC}$ reconstructs the file by connecting to any $k$ storage nodes through edges with infinite capacities. Furthermore, a cut is defined as a set of edges whose removal separates $\mathsf{S}$ from $\mathsf{DC}$, and the minimum cut is the cut with the smallest total capacity among all such separating cuts.

    Now, consider a distributed storage system with parameters $(n,k,d) = (4,2,3)$, storage $\alpha=3$, and per-helper bandwidth $\beta_{\mathsf{c}} = \beta_{\mathsf{q}} \defeq \beta = 1$. The classical cutset bound~(\ref{eq:classical}) gives
    \begin{align}
        \sum_{i=0}^{k-1} \min\{(3-i),3\} = 3 + 2 = 5,
    \end{align}
    which corresponds to the cut illustrated in Fig.~\ref{fig:thm1}. Hence, the system can store a file of size at most $B=5$ dits. In contrast, with quantum communication and entanglement, the corresponding bound in (\ref{eq:main}) becomes
    \begin{align}
        \sum_{i=0}^{1} \min\{2(3-i),3,3\} = 3 + 3 = 6,
    \end{align}
    which corresponds to the same cut in Fig.~\ref{fig:thm1} with increased effective capacity. 
    
    This gain arises because, while the classical cut consists of one storage edge of capacity $\alpha=3$ and two repair edges each of capacity $\beta=1$, the quantum communication links can jointly convey additional information. In particular, due to the superdense coding gain, two quantum links carrying a total of $2\beta$ qudits can transmit up to $4 \beta$ dits of classical information \cite{nielsen2010quantum}, while the Holevo bound \cite{holevo1973bounds} requires that the total amount of information is restricted by the total number of transmitted qudits from the $3$ helper nodes, i.e., $3 \beta$. Hence, the two quantum links can send $\min\{4\beta, 3\beta\} = 3$ dits. As a result, the entanglement-assisted system provides a strictly larger cut capacity than its classical counterpart.
\end{remark}

From Theorem~\ref{thm_cutset}, the quantum MSR (QMSR) point and the quantum MBR (QMBR) point can be derived explicitly in the following corollaries.

\begin{corollary}\label{cor:QMSR}
    The QMSR point is given by the pair
    \begin{align}
         &\left(\alpha^{\mathrm{QMSR}},d \beta_{\mathsf{q}}^{\mathrm{QMSR}} \right)\notag \\
         & \qquad= \left( \frac{B}{k},  \frac{B}{k}  \max \left\{ 1, \frac{d}{2(d-k+1)} \right\} \right) \\
         & \qquad=
            \begin{cases}
                \left(
                \dfrac{B}{k},\,
                \dfrac{B}{k}
                \right),
                & \text{if } d \geq 2k-2, \\
                \left(
                \dfrac{B}{k},\,
                \dfrac{Bd}{2k(d-k+1)}
                \right),
                & \text{if } k \leq d < 2k-2 .
            \end{cases}
    \end{align}
\end{corollary}

\begin{figure}[t]
    \centering
    \includegraphics[width=0.47\textwidth]{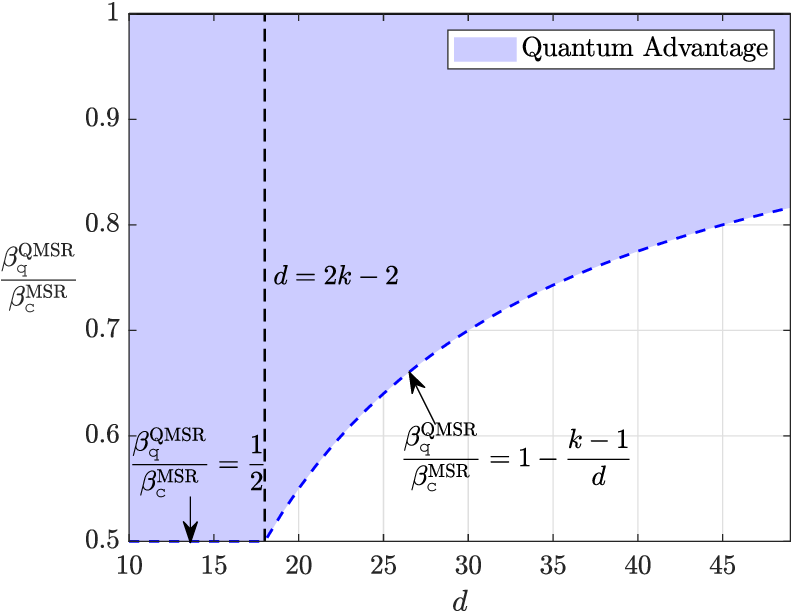}
    \caption{The per-node bandwidth ratio $\frac{\beta_{\mathsf{q}}^{\rm QMSR}}{\beta_{\mathsf{c}}^{\rm MSR}}$ versus the number of helper nodes $d$, when $n=50$, $k=10$. }
    \label{fig:QMSR}
\end{figure}

\begin{remark}
    Compared with the classical MSR point, the minimum storage remains unchanged, i.e., $\alpha^{\rm MSR} = \alpha^{\rm QMSR} = \frac{B}{k}$. However, the required bandwidth with quantum repair is strictly smaller. In particular, the ratio between the quantum and classical per-helper bandwidths is given by
    \begin{align}
        \frac{\beta_{\mathsf{q}}^{\rm QMSR}}{\beta_{\mathsf{c}}^{\rm MSR}} 
        & = \max\left\{1 - \frac{k-1}{d}, \frac{1}{2}\right\}\\
        & = \begin{cases}
                1 - \dfrac{k-1}{d}, & \text{if } d \ge 2k-2,\\
                \dfrac{1}{2},   & \text{if } k \leq d < 2k-2.
            \end{cases}
    \end{align}
    This ratio is illustrated in Fig.~\ref{fig:QMSR}. Observe that the bandwidth advantage enabled by quantum resources is maximized in the regime of small repair degree, i.e., $d \leq 2k-2$, where the quantum bandwidth is exactly one half of its classical counterpart. As $d$ increases beyond this threshold, the relative gain gradually diminishes and vanishes asymptotically as $d \rightarrow \infty$. This is because, for large repair degrees $d$, the classical repair process already approaches its optimal efficiency, i.e., $\frac{d\beta_{\mathsf{c}}^{\rm MSR}}{\alpha^{\rm MSR}} \rightarrow 1$, leaving limited room for further improvement via quantum communication.
\end{remark}

\begin{corollary}\label{cor:QMBR}
    The QMBR point is given by the pair
    \begin{align}
        &\left( \alpha^{\rm QMBR}, d\beta_{\mathsf{q}}^{\rm QMBR} \right) \notag\\
        &\qquad=
        \begin{cases}
        \left(
        \dfrac{B}{k},\, \dfrac{B}{k}
        \right),
        & \text{if } d \ge 2k-2, \\
        \left(
        \bar{\alpha},\, \bar{\alpha}
        \right),
        & \text{if } k \le d < 2k-2,
        \end{cases}
        \end{align}
        where
        \begin{align}
        \bar{\alpha}
        \defeq
        \frac{dB}{
        d\bigl(\lfloor d/2 \rfloor + 1\bigr)
        +
        \bigl(2d - k - \lfloor d/2 \rfloor\bigr)
        \bigl(k - \lfloor d/2 \rfloor - 1\bigr)
        }.
    \end{align}
\end{corollary}

\begin{figure}[t]
      \centering
      \includegraphics[width=0.47\textwidth]{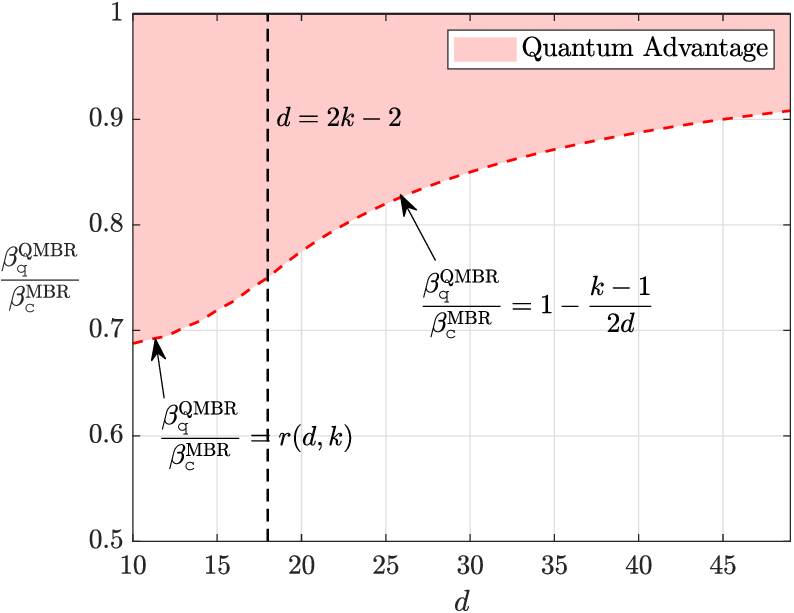}
      \caption{The per-node bandwidth ratio $\frac{\beta_{\mathsf{q}}^{\rm QMBR}}{\beta_{\mathsf{c}}^{\rm MBR}}$ versus the number of helper nodes $d$, when $n=50, k=10$. }
      \label{fig:QMBR}
\end{figure}

\begin{remark}
    We now compare the QMBR point with the classical MBR point. Since both points satisfy $\alpha = d \beta$, it suffices to compare the ratio of the per-helper repair bandwidths $\frac{\beta_{\mathsf{q}}^{\rm QMBR}}{\beta_{\mathsf{c}}^{\rm MBR}}$.
    \begin{itemize}
        \item When $ d\geq 2k-2$, the ratio between the quantum and classical per-helper repair bandwidths is given by
        \begin{align}
            \frac{\beta_{\mathsf{q}}^{\rm QMBR}}{\beta_{\mathsf{c}}^{\rm MBR}} = 1-\frac{k-1}{2d}.
        \end{align}
        Hence,
        \begin{align}
            \frac{3}{4}  \leq \frac{\beta_{\mathsf{q}}^{\rm QMBR}}{\beta_{\mathsf{c}}^{\rm MBR}} \leq 1.
        \end{align}
        Moreover, for a fixed $k$, this ratio is a monotonically increasing function of $d$, rising from $3/4$ and approaching $1$ as $d$ grows. This indicates that the relative bandwidth reduction offered by quantum communication becomes less significant when the repair degree is much larger than $k$.
        \item When $ k \leq d\leq 2k-2$, the ratio between the quantum and classical per-helper bandwidth is
        \begin{align}
            & r(d,k) \defeq \frac{\beta_{\mathsf{q}}^{\rm QMBR}}{\beta_{\mathsf{c}}^{\rm MBR}  } \notag \\
            =& \frac{k(2d-k+1)}{2d (\lfloor d/2 \rfloor + 1) + 2(2d - k - \lfloor d/2 \rfloor)(k - \lfloor d/2 \rfloor - 1)},
        \end{align}
        which depends on whether $d$ is odd or even. Nevertheless, for a fixed $k$, the ratio $r(d,k)$ increases monotonically with $d$ over the interval $k \leq d \leq 2k-2$, and it reaches $3/4$ when $d = 2k-2$.
    \end{itemize}
    Overall, the per-node bandwidth ratio $\frac{\beta_{\mathsf{q}}^{\rm QMBR}}{\beta_{\mathsf{c}}^{\rm MBR}  }$ is a monotonically increasing function of the number of helper nodes $d$ over the entire range $k \leq d \leq n-1$. In particular, when $k \leq d \leq 2k-2$, the ratio increases with $d$ and attains the value $3/4$ at $d=2k-2$. For $d \geq 2k-2$, the ratio continues to increase from $3/4$ and asymptotically approaches $1$ as $d$ grows. The curve is illustrated in Fig.~\ref{fig:QMBR}.
\end{remark}

\begin{remark}[\textbf{Optimal regenerating point}]
    By comparing Corollary \ref{cor:QMSR} and Corollary \ref{cor:QMBR}, we observe that the QMSR and QMBR points coincide when $d \geq 2k-2$. In this regime, there is a single optimal regenerating point that simultaneously minimizes both the storage cost and the repair bandwidth, thereby breaking the fundamental storage–bandwidth tradeoff present in classical systems.
    This phenomenon arises because, when $d \geq 2k-2$, entanglement assistance enables the repair bandwidth at the QMSR point to attain its minimum value, $d \beta_{\mathsf{q}}^{\rm QMSR} = \alpha^{\rm QMSR} $, which also achieves the minimization of the bandwidth, i.e., the QMBR point. Consequently, the QMSR and QMBR points coincide.
\end{remark}

\section{Illustrative Examples}

\begin{figure}[t]
    \centering
    \includegraphics[width=0.48\textwidth]{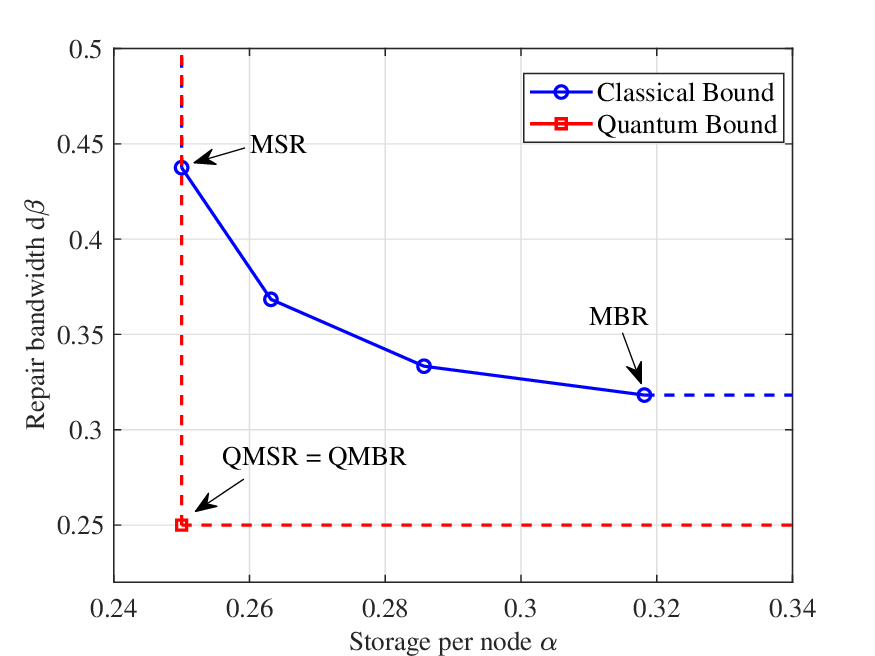}
    \caption{Comparison of the classical and quantum tradeoff for $n=8$, $k = 4$, $d = 7$, $B=1$. Since $d \geq 2k-2$, the QMSR point coincides with the QMBR point, yielding a single optimal regenerating point for the system.}
    \label{fig:ex1}
\end{figure}

\begin{example}\label{example1} ($d \geq 2k-2$)
    Consider a distributed storage system with parameters $(n,k,d)=(8,4,7)$, and normalize the file size $B=1$. We now compare the storage-bandwidth tradeoff in classical and entanglement-assisted settings. From (\ref{eq:classical}), the classical storage--bandwidth tradeoff is given by
    \begin{align}
        \sum_{i=0}^{3} \min\{(7-i) \beta_{\mathsf{c}}, \alpha \} \geq 1.
    \end{align}
    The corresponding MSR point is thus
    \begin{align}
       (\alpha^{\rm MSR},d\beta_{\mathsf{c}}^{\rm MSR}) = \left(\frac{1}{4}, \frac{7}{16} \right),
    \end{align}
    while the MBR point is given by
    \begin{align}
        (\alpha^{\rm MBR},d\beta_{\mathsf{c}}^{\rm MBR}) = \left(\frac{7}{22}, \frac{7}{22} \right),
    \end{align}
    where the repair bandwidth equals the per-node storage at the MBR point.
    
    In contrast, under entanglement assistance, Theorem~\ref{thm_cutset} yields the bound
    \begin{align}
        \sum_{i=0}^{3} \min\{2(7-i)\beta_{\mathsf{q}},7\beta_{\mathsf{q}},\alpha\} = 4\min\{ \alpha, 7\beta_{\mathsf{q}} \} \geq 1,
    \end{align}
    as shown in (\ref{eq:main}). Hence, the QMSR and QMBR operating points are given by
    \begin{align}
        (\alpha^{\rm QMSR},d\beta_{\mathsf{q}}^{\rm QMSR}) = (\alpha^{\rm QMBR},d\beta_{\mathsf{q}}^{\rm QMBR}) = \left(\frac{1}{4},\frac{1}{4} \right).
    \end{align}
    The classical and quantum curves are shown in Fig.~\ref{fig:ex1}. It can be seen that, in the case of minimum storage, we have $\alpha^{\rm QMSR} = \alpha^{\rm MSR}$, but the bandwidth is significantly reduced. Indeed, when $d \geq 2k-2$, entanglement assistance allows the repair bandwidth at the QMSR point to attain its minimum possible value, equal to the per-node storage.
    This also leads to the minimum bandwidth point. Hence, the QMSR and QMBR points coincide, yielding a single optimal point on the quantum curve.
\end{example}

\begin{figure}[t]
    \centering
    \includegraphics[width=0.48\textwidth]{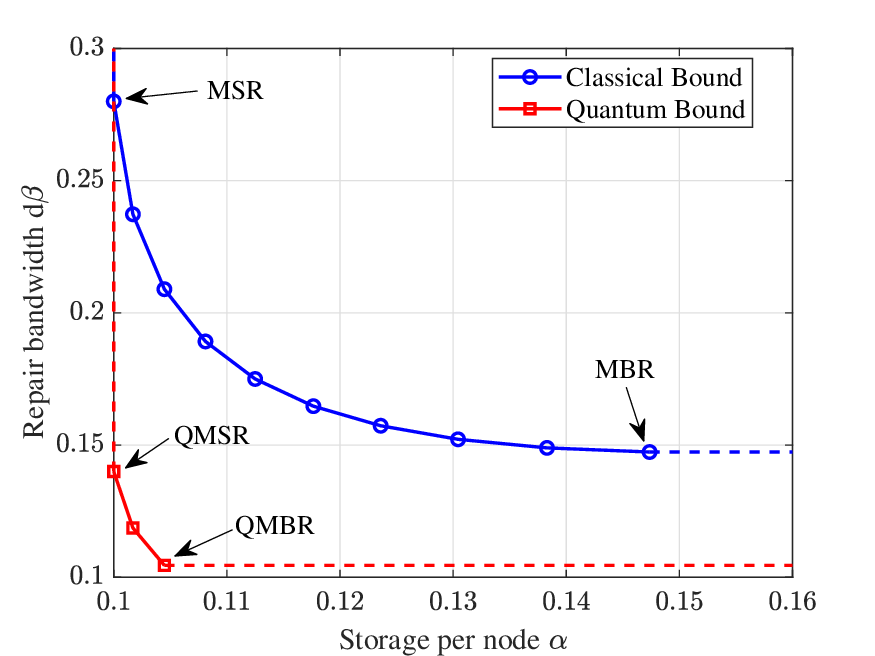}
    \caption{Comparison of the classical and quantum tradeoff curves for $n=15$, $k=10$, $d=14$, and $B=1$.}
    \label{fig:ex2}
\end{figure}

\begin{example}\label{example2}($d < 2k-2$) 
    In this example, we focus on the other regime when the number of helper nodes is relatively small, i.e., $d < 2k-2$. Consider a distributed storage system with parameters $(n,k,d)=(15,10,14)$ and a normalized file size $B=1$. In the classical setting, the storage–bandwidth tradeoff is characterized by
    \begin{align}
        \sum_{i=0}^{9} \min \{ (14-i) \beta_{\mathsf{c}}, \alpha\} \geq 1.
    \end{align}
    The corresponding MSR and MBR points are given by
    \begin{align}
        (\alpha^{\rm MSR},d\beta_{\mathsf{c}}^{\rm MSR}) & = \left(\frac{1}{10}, \frac{7}{25}  \right), \\ (\alpha^{\rm MBR},d\beta_{\mathsf{c}}^{\rm MBR}) & = \left(\frac{14}{95}, \frac{14}{95}  \right),
    \end{align}
    respectively.
    
    In the quantum setting, the  bound in Theorem~\ref{thm_cutset} specializes to
    \begin{align}
        \sum_{i=0}^{9} \min \{14 \beta_{\mathsf{q}}, 2(14-i) \beta_{\mathsf{q}}, \alpha\} \geq 1,
    \end{align}
    which yields the QMSR and QMBR points
    \begin{align}
        (\alpha^{\rm QMSR},d\beta_{\mathsf{q}}^{\rm QMSR}) & = \left(\frac{1}{10}, \frac{7}{50}  \right), \\ (\alpha^{\rm QMBR},d\beta_{\mathsf{q}}^{\rm QMBR}) & = \left(\frac{7}{67}, \frac{7}{67}  \right),
    \end{align}
    respectively. The classical and quantum curves are shown in Fig.~\ref{fig:ex2}.
    
    Note that, unlike the regime $d \geq 2k-2$, there is no single optimal point in the quantum setting. However, it can be observed that, the quantum tradeoff outperforms its classical counterpart across the entire curve.
    In particular, entanglement-assisted repair achieves a substantial reduction in repair bandwidth at the minimum-storage point. Recalling Corollary~2, since $d \leq 2k-2$ in this example, the repair bandwidth at the QMSR point is reduced by a factor of two compared to the classical MSR point.
\end{example}

\newpage

\bibliographystyle{unsrt}
\bibliography{Ref1.bib}

\end{document}